\documentclass[twocolumn,floats,prb,aps,showpacs]{revtex4}

\usepackage[dvips]{graphicx}

\begin{document}

\title{Infrared Signature of the
       Superconducting State in Pr$_{2-x}$Ce$_x$CuO$_4$}

\author{A. Zimmers}
\author{R.P.S.M. Lobo}
\email{lobo@espci.fr}
\author{N. Bontemps}
\affiliation{Laboratoire de Physique du Solide (UPR 5 CNRS) ESPCI,
10 rue Vauquelin 75231 Paris, France.}

\author{C.C. Homes}
\affiliation{Department of Physics, Brookhaven National Laboratory, Upton, 
    New York 11973, USA.}

\author{M.C. Barr}
\author{Y. Dagan}
\author{R.L. Greene}
\affiliation{Center for Superconductivity Research, Department of Physics, 
    University of Maryland, College Park, Maryland 20742, USA.}

\date{\today}

\begin{abstract} 
We measured the far infrared reflectivity of two 
superconducting Pr$_{2-x}$Ce$_x$CuO$_4$ films above and below $T_c$. The 
reflectivity in the superconducting state increases and the optical conductivity 
drops at low energies, in agreement with the opening of a (possibly) anisotropic 
superconducting gap. The maximum energy of the gap scales roughly with $T_c$ as 
$2 \Delta_\textrm{max} / k_B T_c \approx 4.7$. We determined absolute values of 
the penetration depth at 5 K as $\lambda_{ab} = (3300 \pm 700) \textrm{ \AA}$ 
for $x = 0.15$ and $\lambda_{ab} = (2000 \pm 300) \textrm{ \AA}$ for $x = 0.17$. 
A spectral weight analysis shows that the Ferrell-Glover-Tinkham sum rule is 
satisfied at conventional low energy scales $\sim 4 \Delta_\textrm{max}$.
\end{abstract}

\pacs{74.25.Gz, 74.72.Jt}

\maketitle


Since the discovery of High-$T_c$ superconductivity, considerable efforts have 
been made to describe and understand the response of the superconducting phase. 
Over the years, measurements have been carried out in larger temperature and 
doping ranges, hoping to gain new insight into the problem.

The discovery of electron-doped cuprates\cite{Tokura,Takagi} gave access to the 
mirror image of the hole doped phase diagram with respect to the Mott insulator 
state. Since then, significant work has been made looking for the differences 
and similarities in systems with either type of carrier.\cite{FournierReview} 
The general phase diagram presents global symmetry, yet the magnetic properties 
show clear differences, the most obvious being the much broader 
antiferromagnetic phase on the electron doped side. 

On the hole-doped side the main results from infrared spectroscopy can be 
summarized as: (i) in the normal state, indirect evidence of the pseudogap phase 
has come from analysis of the inverse quasiparticle lifetime $1/\tau(\omega)$, 
which is depleted over a range $\approx 100~meV$;\cite{Timusk} (ii) the system 
evolves smoothly from the normal state into the superconducting state with no 
typical energy scaling with $T_c$;\cite{Orenstein} (iii) for most dopings these 
materials are in the clean limit;\cite{Puchkov} and (iv) a non conventional 
pairing mechanism is supported by some evidence that high energy states 
contribute to the formation of the condensate in the underdoped 
regime.\cite{Santander,Molegraaf,Homes1}

Recent studies above $T_c$, in the electron-doped side, have concluded through 
direct spectral weight analysis that a high energy partial gap opened in the 
normal state.\cite{Zimmers, Onose1, Onose2} However very little is known about 
the optical properties in the superconducting state, most likely due to the low 
energy associated with the superconducting gap.\cite{Homes2,Lupi,Singley2001} 
Indeed, most studies rely on the Raman scattering of Nd$_{2-x}$Ce$_x$CuO$_4$ 
(NCCO) and Pr$_{2-x}$Ce$_x$CuO$_4$ (PCCO) at optimal doping (maximum $T_c$)
$x=0.15$.\cite{Kendziora1, Kendziora2, Blumberg}

In this paper, we take advantage of the large surface and good homogeneity of 
Pr$_{2-x}$Ce$_x$CuO$_4$ films to explore the changes induced by 
superconductivity in the far-infrared spectra. Our data shows an enhanced 
reflectivity at low frequencies when going into the superconducting state, on an 
energy scale entirely different from that of the normal state 
gap.\cite{Zimmers, Onose1, Onose2} This feature is translated as a spectral 
weight loss in the real part of the optical conductivity. Comparing one 
overdoped to one optimally doped sample we show that the energy scale associated 
with the superconducting gap roughly scales with $T_c$.


Two Pr$_{2-x}$Ce$_x$CuO$_4$ films were epitaxially grown by pulsed-laser 
deposition on SrTiO$_3$ and annealed in reducing 
atmosphere.\cite{FournierSample} The optimally doped sample is obtained with 
$x = 0.15$, has $T_c = 21$~K and is 3780~\AA \mbox{ } thick. The sample with 
$x = 0.17$ is in the overdoped regime, has $T_c = 15$~K and a thickness of 
3750~\AA. The critical temperatures were obtained by electrical transport and 
are defined by the zero resistance. Thin films are extremely homogeneous in the 
Ce concentration and their large surface to volume ratio makes them easy to 
anneal.

Near normal incidence infrared and visible reflectivity spectra were taken 
between $60$ and $21000 \textrm{ cm}^{-1}$ in a Bruker IFS 66v interferometer at 
ESPCI. This data was extended to the very far-infrared 
($10$--$200\textrm{ cm}^{-1}$) at Brookhaven National Laboratory utilizing a 
Bruker IFS 113v interferometer. The films had an exposed area to the infrared 
light of about 5~mm in diameter. At ESPCI, gold mirrors were used as a reference 
below $10000\textrm{ cm}^{-1}$ and silver mirrors above $8000\textrm{ cm}^{-1}$. 
The data is corrected for the absolute reflectivity of Au and Ag. The 
reflectivity measured at BNL uses an in-situ overcoating of the sample by gold 
as a reference.\cite{Homes3} Measures taken at ESPCI and at BNL agree within 
0.2~\%, setting our estimate of error in the absolute reflectivity. The relative 
accuracy of the measurement (between two consecutive temperatures, for instance) 
is estimated to be better than 0.1~\%.

Data was taken at several temperatures in the whole spectral range but in this 
paper we are only going to compare the far-infrared spectrum (below 
$500\textrm{ cm}^{-1}$) just above $T_c$ to the one at 5~K. The normal state 
properties and the full spectral range are discussed elsewhere.\cite{Zimmers}

The high reflectivity of the film below $200\textrm{ cm}^{-1}$ makes the 
substrate contribution negligible in this range. Substrate contributions to the 
reflectivity cannot however be neglected above $200\textrm{ cm}^{-1}$. To get 
rid of this response, we measured the reflectivity of SrTiO$_3$ at the same 
temperatures in order to extract its optical properties. We then searched for a 
dielectric function for the film that describes the reflectivity of the whole 
system using a standard thin film model.\cite{Santander1} The dielectric function
obtained for the film was used to generate its bulk reflectivity above 
$100\textrm{ cm}^{-1}$ at all temperatures. The reflectivity spectra beyond 
$100\textrm{ cm}^{-1}$ are identical within the accuracy of the measurement, 
from 25~K, just above $T_c$, down to the lowest temperature. 

The overall bulk reflectivity of PCCO can then be obtained by combining the 
measured data below $200\textrm{ cm}^{-1}$ to the bulk simulation above 
$100\textrm{ cm}^{-1}$.\cite{Overlap} Finally, we applied standard 
Kramers-Kronig analysis to such reconstructed spectra in order to extract the 
optical conductivity of PCCO. Below $20\textrm{ cm}^{-1}$ we used a Hagen-Rubens 
($1 - a \sqrt{\omega}$) extrapolation for the normal state reflectivity and a 
superconductor extrapolation ($1 - b \omega^4$) below $T_c$. Above 
$21000 \textrm{ cm}^{-1}$ we used a constant up to $10^6 \textrm{ cm}^{-1}$ 
followed by a free electron $1 / \omega^4$ termination.

\begin{figure}
  \begin{center}
    \includegraphics[width=8cm]{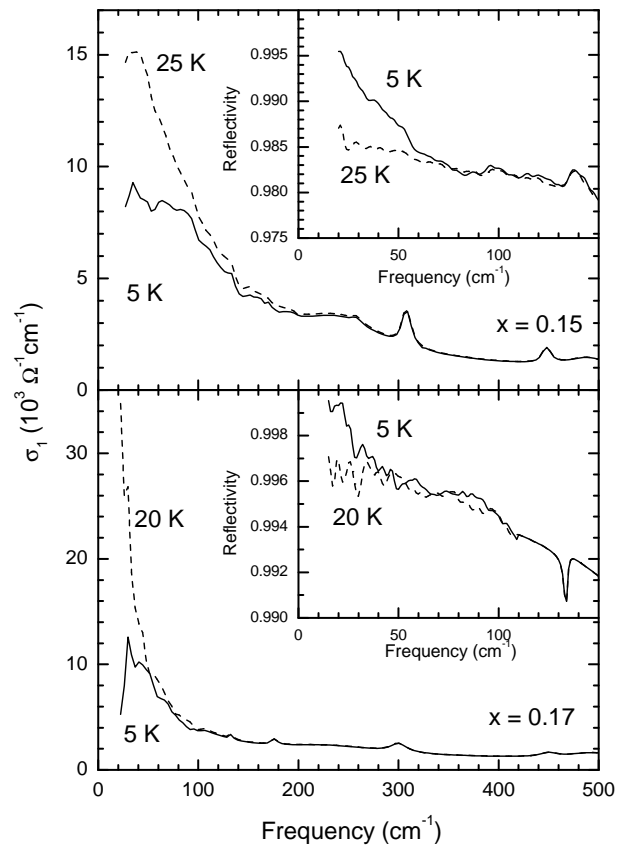}
  \end{center}
\caption{Real part of the optical conductivity for the optimally (top panel) and 
overdoped (bottom panel) Pr$_{2-x}$Ce$_x$CuO$_4$. The insets show the 
far-infrared reflectivity for the respective samples. In all panels the dashed 
line corresponds to a temperature just above $T_c$ and the solid line to 5~K.}
\label{fig1}
\end{figure}

Figure \ref{fig1} shows the real part of the optical conductivity ($\sigma_1$) 
for $x = 0.15$ (top panel) and $0.17$ (bottom panel). The insets in this figure 
show the measured far-infrared reflectivity. In all panels the dashed line is 
taken just above $T_c$ and the solid line at 5~K. In both compounds the far 
infrared reflectivity increases, corresponding to a depletion in the 
superconducting $\sigma_1$ at low frequencies.

\begin{figure}
  \begin{center}
    \includegraphics[width=8cm]{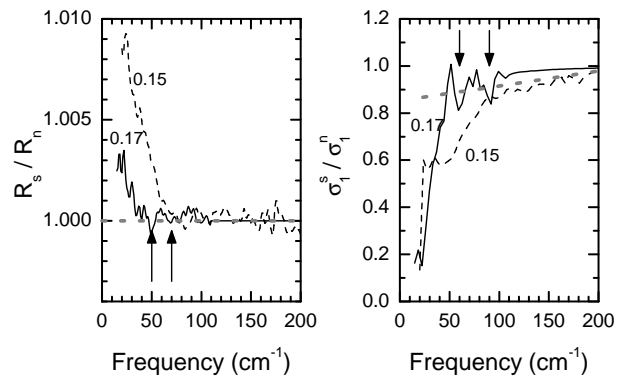}
  \end{center}
\caption{Superconducting to normal state ratio of the reflectivity (left panel) 
and optical conductivity (right panel) in PCCO. The dashed line is for 
$x = 0.15$ and the solid line for $x = 0.17$. The dotted straight lines are 
guides for the eye representing the average high frequency behavior. The arrows 
indicate the frequency where this linear behavior breaks down.}
\label{fig2}
\end{figure}

In Fig. \ref{fig2} we plot the ratio between superconducting and normal 
reflectivity (left panel) and conductivity (right panel) for the $x = 0.15$ 
(dashed line) and $x = 0.17$ (solid line) samples. We note that there is an 
increase in the low energy reflectivity at $70 \textrm{ cm}^{-1}$ for x = 0.15 
and $50 \textrm{ cm}^{-1}$ for $x = 0.17$. The corresponding decreases in the 
optical conductivity occurs at $90 \textrm{ cm}^{-1}$ and 
$60 \textrm{ cm}^{-1}$.

In a $s$-wave BCS superconductor a rise in the low frequency reflectivity is 
associated with an isotropic superconducting gap ($2 \Delta$). However the BCS 
reflectivity is much flatter and closer to unity than what is seen in our data. 
Nevertheless, the reflectivity rise is compatible with the onset of an 
anisotropic gap. Indeed, the two strongest arguments against the observation of 
the gap in cuprates are (i) the energy range where the reflectivity increases 
does not vary with doping and (ii) cuprates are thought to be in the clean limit 
making the observation of a gap difficult. The first argument is clearly not 
applicable to our data. To counter the second point we can look at the low
frequency scattering rate just above $T_c$. In the optimally doped sample we 
have $1 / \tau(0) \approx 85 \textrm{ cm}^{-1}$ and in the overdoped material 
$1 / \tau(0) \approx 30 \textrm{ cm}^{-1}$. These values are of the same order 
of the frequency where the reflectivity increases. It is then reasonable to 
assign the reflectivity rise and the conductivity drop to the superconducting 
gap. In the absence of a specific model for such a gap, we can only estimate the 
maximum gap value from the frequencies where the low energy reflectivity or 
conductivity in the superconducting state differs from the ones above $T_c$. If 
we use the values obtained from $\sigma_1$ we have a 
$2 \Delta_\textrm{max} / k_B T_c$ ratio of 6 for $x = 0.15$ and 5.6 for 
$x = 0.17$. This value is probably an overestimate of the gap energy. 
Considering the frequencies obtained from the reflectivity, the 
$2 \Delta_\textrm{max} / k_B T_c$ ratio is 4.7 for both samples, in closer 
agreement to the values inferred from the Raman $B_{2g}$ symmetry in NCCO 
samples.\cite{Kendziora1, Kendziora2, Blumberg}

To further quantify the superconducting properties of PCCO, we looked at the 
$f$-sum rule. It follows from charge conservation and states that
\begin{equation}
  \int_0^\infty \sigma_1(\omega^\prime) d\omega^\prime =
    \frac{\pi^2}{Z_0} \frac{n e^2}{m},
  \label{eq1}
\end{equation}
where $Z_0 \approx 377~\Omega$ is the vacuum impedance, $n$ the charge density, 
and $e$ and $m$ the electronic charge and mass respectively.

Infinite conductivity in the superconducting state is represented by a 
$\delta(\omega)$ peak at the origin in $\sigma_1$. The $f$-sum rule then implies 
that the spectral weight of the $\delta(\omega)$ peak must come from finite 
frequencies, hence the decrease in $\sigma_1$. In fact, Ferrell, Glover and 
Tinkham (FGT)\cite{FGT1,FGT2} have shown that the spectral weight lost at 
finites frequencies of $\sigma_1$ in the superconducting state is recovered in 
the superfluid weight. 

To verify the FGT sum rule we need to determine the superfluid weight. One way 
to calculate it is to use the imaginary part $\sigma_2$ of the optical 
conductivity.\cite{Dordevic} In the superconducting state one can write 
$\sigma_1$ as
\begin{equation}
  \sigma_1(\omega) = \frac{\pi^2}{Z_0} \Omega_{sc}^2 \delta(\omega) +
  					 \sigma_1^\prime(\omega),
  \label{eq2}
\end{equation}
where $\Omega_{sc}$ is the superconducting plasma frequency. $\sigma_2$ then 
follows from Kramers-Kronig as
\begin{equation}
  \sigma_2(\omega) = \frac{2\pi}{Z_0 \omega} \Omega_{sc}^2 -
  					 \frac{2 \omega}{\pi}
  					 \int_0^\infty
  					 \frac{\sigma_1^\prime(\omega^\prime)}
  					      {{\omega^\prime}^2 - \omega^2} d \omega^\prime
  \label{eq3}
\end{equation}
and we define $\sigma_2^{sc}$ as the first term on the right hand side of 
Eq. \ref{eq3}.

Kramers-Kronig of the reflectivity data yields directly $\sigma_1$ and 
$\sigma_2$. Because $\sigma_1$ is obtained only for finite frequencies, it 
equals $\sigma_1^\prime$. We can thus apply Eq. \ref{eq3} and calculate 
$\sigma_2^{sc}$.

\begin{figure}
  \begin{center}
    \includegraphics[width=8cm]{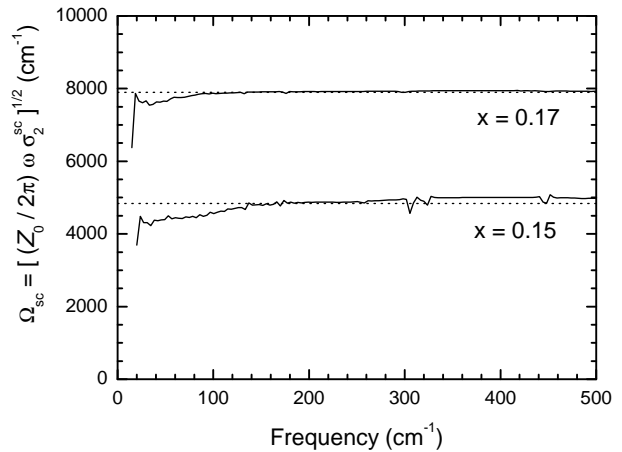}
  \end{center}
\caption{Superconducting plasma frequency extracted from the $1 / \omega$ 
component of $\sigma_2$ in the superconducting state. The dotted lines are the 
average values for $\Omega_{sc}$ between $20$ and $500 \textrm{ cm}^{-1}$.}
\label{fig3}
\end{figure}

Figure \ref{fig3} shows that $\sqrt{\omega \sigma_2^{sc}}$ for both samples is 
indeed fairly constant below $500 \textrm{ cm}^{-1}$ and corresponds to 
$\Omega_{sc} = (4800 \pm 1000) \textrm{ cm}^{-1}$ for $x = 0.15$ and 
$\Omega_{sc} = (7900 \pm 750) \textrm{ cm}^{-1}$ for $x = 0.17$. The 
contributions to errors in $\Omega_{sc}$ come from (i) the uncertainties in 
fitting the low frequency $\omega \sigma_2^{sc}$ to a constant value; and (ii) 
different extrapolations used in the Kramers-Kronig calculations. The absolute 
value of the superconducting penetration depth at 5~K can be calculated using 
$\lambda_{ab}=\frac{1}{2\pi}{\Omega_{sc}}^{-1}$ and  yield 
$\lambda_{ab} = (3300 \pm 700) \textrm{ \AA}$ for $x = 0.15$ and 
$\lambda_{ab} = (2000 \pm 300) \textrm{ \AA}$ for $x = 0.17$. These values are 
in good agreement with the ones obtained by microwave 
absorption.\cite{Kokales,Snezhko}

We can now verify the FGT sum rule by comparing the spectral weight lost in 
$\sigma_1$ to the superconducting plasma frequency calculated above. To do so, 
we define a partial differential sum rule
\begin{equation}
  \Lambda^2(\omega) = \frac{Z_0}{\pi^2}
  	\int_{0^+}^\omega [ \sigma_1^N(\omega^\prime) -
  	                \sigma_1^S(\omega^\prime)] d\omega^\prime,
  \label{eq4}
\end{equation}
where the superscripts $N$ and $S$ refer to the normal and superconducting 
states respectively. One should read $\Lambda(\omega)$ as the contribution from 
states up to $\omega$ to the the superconducting plasma frequency, {\it i.e.},
$\Omega_{sc} = \Lambda(\omega \rightarrow \infty)$.

\begin{figure}
  \begin{center}
    \includegraphics[width=8cm]{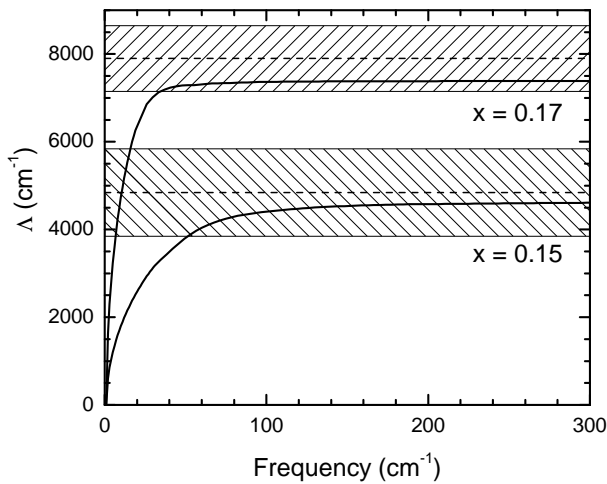}
  \end{center}
\caption{Partial differential sum rule for optimally and overdoped PCCO 
calculated using Equation \ref{eq4}. The dashed lines are the superconducting 
plasma frequencies calculated from Fig. \ref{fig3}. The shaded areas indicate 
the error in these values.}
\label{fig4}
\end{figure}

Figure \ref{fig4} shows $\Lambda(\omega)$ for both films. To correctly obtain 
$\Lambda(\omega)$, one must integrate from $0^+$ but our data only goes down to 
$20 \textrm{ cm}^{-1}$. To get the area between $0$ and $20 \textrm{ cm}^{-1}$, 
we fitted the normal state $\sigma_1$ using a Drude peak. For lack of a good 
extrapolation, the superconducting $\sigma_1$ was set to $0$ below 
$20 \textrm{ cm}^{-1}$. Inspection of Fig. \ref{fig1} shows that this 
approximation is reasonable for the overdoped sample. However it will largely 
overestimate $\Lambda(\omega)$ in the optimally doped compound. In that case we 
used a constant extrapolation to zero frequency. The dashed lines in 
Fig. \ref{fig4} are the superconducting plasma frequencies calculated from 
Fig. \ref{fig3}. The shaded areas indicate the error in these values.

Figure \ref{fig4} is showing how far one must integrate the conductivity to 
obtain $\Omega_{sc}$ from $\Lambda(\omega)$. Within error bars the FGT sum rule 
is fulfilled in both samples below $200 \textrm{ cm}^{-1}$ (25 meV). A similar 
analysis in underdoped Bi$_2$Sr$_2$CaCu$_2$O$_{8-\delta}$ shows that one must 
integrate Eq. \ref{eq4} to very high energies (2 eV) in order to recover the 
spectral weight of the condensate.\cite{Santander,Molegraaf} A similar effect 
was also seen in YBa$_2$Cu$_3$O$_{6.5}$ where the integration must be carried up 
to 0.5 eV to satisfy the FGT sum rule.\cite{Homes1} This energy range is 
characteristic of the boson spectrum responsible for the pairing mechanism which 
led to the conclusion that the pairing mechanism was 
unconventional.\cite{Norman, Hirsch, Benfatto, Stanescu, Ashkenazi, Carbotte}
In our PCCO samples, Fig. \ref{fig4} shows that the states contributing to the 
formation of the superconducting plasma frequency lie at energies comparable to 
the phonon spectrum.

The energy scale over which the condensate is recovered as well as the 
superconducting gap value are much smaller than the magnitude of the normal 
state (pseudo) gap observed around 100 meV.\cite{Onose1, Onose2,Zimmers} This is 
in striking contrast with hole doped cuprates where the area loss in the real 
part of the conductivity, due to  superconductivity, occurs over an energy scale 
which is similar to the one associated with the pseudogap state. This might 
imply that the normal state gap in electron doped cuprates has a different 
microscopic origin from the pseudogap in hole doped cuprates.


In conclusion, we have measured the optical conductivity of one optimally doped 
and one overdoped Pr$_{2-x}$Ce$_x$CuO$_4$ film. The reflectivity increase at low 
frequencies can be associated with the superconducting gap maximum and its value 
scales with $T_c$ as $2\Delta_\textrm{max} \approx 4.7 k_B T_c$. The superconducting 
penetration depth at 5 K was determined to be 
$\lambda_{ab} = (3300 \pm 700) \textrm{ \AA}$ for the optimally doped sample and 
$\lambda_{ab} = (2000 \pm 300) \textrm{ \AA}$ for the overdoped one. The partial 
differential sum rule shows that the superfluid condensate is built from states 
below 25 meV compatible with a more conventional low energy pairing mechanism.

The authors thank Dr. V.N. Kulkarni for RBS / Channeling measurements and 
C.P. Hill for help in the sample preparation. The work at University of Maryland 
was supported by NSF grant DMR-0102350. The Work at Brookhaven National 
Laboratory was supported by DOE under contract DE-AC02-98CH10886.


\end{document}